\documentclass[aps,prl,twocolumn,showpacs,reprint,superscriptaddress]{revtex4}
\usepackage{epsfig,amsmath,amssymb,graphics,color,calc}
\usepackage[ansinew]{inputenc}

\newcommand{\be}{\begin{equation}}
\newcommand{\ee}{\end{equation}}
\newcommand{\ba}{\begin{eqnarray}}
\newcommand{\ea}{\end{eqnarray}}

\renewcommand{\phi}{\varphi}
\newcommand{\taue} {\tau_{1/\mathrm{e}}}
\newcommand{\edot} {\dot{\epsilon}}
\newcommand{\micron} {~\mu\mathrm{m}}

\begin{document}

\title{Highly nonlinear dynamics in a slowly sedimenting colloidal gel}

\author{G. Brambilla}
\affiliation{Universit\'{e} Montpellier 2, Laboratoire Charles
Coulomb UMR 5221, F-34095, Montpellier,
France\\
CNRS, Laboratoire Charles Coulomb UMR 5221, F-34095, Montpellier,
France}

\author{S. Buzzaccaro}
\affiliation{Dipartimento di Chimica, Politecnico di Milano, 20131
Milano, Italy}

\author{R. Piazza}
\affiliation{Dipartimento di Chimica, Politecnico di Milano, 20131
Milano, Italy}

\author{L. Berthier}
\affiliation{Universit\'{e} Montpellier 2, Laboratoire Charles
Coulomb UMR 5221, F-34095, Montpellier,
France\\
CNRS, Laboratoire Charles Coulomb UMR 5221, F-34095, Montpellier,
France}

\author{L. Cipelletti}
\affiliation{Universit\'{e} Montpellier 2, Laboratoire Charles
Coulomb UMR 5221, F-34095, Montpellier,
France\\
CNRS, Laboratoire Charles Coulomb UMR 5221, F-34095, Montpellier,
France} \email{luca.cipelletti@univ-montp2.fr}

\date{\today}



\begin{abstract}
We use a combination of original light scattering techniques and particles with unique optical
properties to investigate the behavior of suspensions of attractive
colloids under gravitational stress, following over time the concentration profile, the velocity profile,
and the microscopic dynamics. During the
compression regime, the sedimentation velocity grows nearly linearly
with height, implying that the gel settling may be fully described
by a (time-dependent) strain rate. We find that the microscopic dynamics
exhibit remarkable scaling properties when time is normalized by
strain rate, showing that the gel microscopic restructuring is
dominated by its macroscopic deformation.
\end{abstract}

\maketitle

Gels and attractive glasses resulting from the aggregation of colloidal particles are the subject of extensive studies
because their physical behavior often results from a complex interplay between
equilibrium thermodynamics and nonequibrium dynamic
processes~\cite{CipellettiJPCM2005,ZaccarelliJPCM2007,royall08},
and because they are relevant for understanding network-forming biological
systems~\cite{kasza07} and for industrial applications.
Although they exhibit solid-like mechanical properties, colloidal gels are easily disrupted by
small perturbations, such as gravitational forces.
While a large body of macroscopic observations of gels under gravitational
stress exists~\cite{BuscallFaradyTrans1987,AllainPRL1995,SenisPRE1997,
PoonFaradayDiscussion1999,StarrsJPCM2002,DerecPRE2003,
ManleyPRL2005Sedimentation,CondreJSTAT2007,KimPRL2007}, very little is
known on the microscopic processes at play during sedimentation, thus limiting
our ability to understand and predict the behavior of sedimenting gels.

Here, we use a novel light scattering method to gain access to the
dynamics of a slowly settling colloidal system from the macroscopic
deformation of the sample down to the relaxational behavior at the
particle scale. We find that the very slow macroscopic deformation
occurs via irreversible plastic events at the microscopic scale.
Remarkably, the gel behavior at all scales is controlled by a single
parameter, the time-dependent compression rate, in striking analogy
with recent observations on deformed polymer~\cite{lee09} and
colloidal~\cite{besseling07} glasses.

We study gels formed by attractive colloidal hard spheres with
radius $R=82 \pm 3$ nm, suspended in an aqueous solvent at an
initial volume fraction $\phi_0 = 0.123$ (more details can be found
in~\cite{Buzzaccaro3,Buzzaccaro,SuppMat}). Gelation is induced by
attractive depletion forces obtained by adding micelles of a
nonionic surfactant. The interaction between colloids is well
described by the Asakura-Oosawa depletion
potential~\cite{Buzzaccaro3}, with a range $r \approx 3$ nm. The
potential can be mapped on the Adhesive Hard Sphere model, with a
stickiness parameter $\tau \simeq 0.01$
~\cite{Buzzaccaro3,Buzzaccaro}. The density mismatch between the
particles and the solvent is $\Delta \rho = 1.12~\mathrm{g/cm^3}$.
The particles have an intrinsic optical anisotropy; accordingly,
they scatter light with polarization both parallel and perpendicular
(``depolarized'') to that of the incident radiation. The depolarized
scattered intensity is an accurate probe of the local particle
concentration~\cite{Buzzaccaro3}.

To probe the sedimentation process in great detail, we use a
custom-designed light scattering apparatus~\cite{DuriPRL2009},
sketched in Fig. SM1~\cite{SuppMat}. A low magnification image of
the sample illuminated by a vertical laser sheet of thickness $200
\micron$ is formed onto a CCD sensor, using depolarized light
scattered at $\theta = 90^{\circ}$. By averaging the CCD signal over
horizontal rows of pixels, we are able to measure accurately the
time, $t$, and height, $z$, dependence of the volume fraction,
$\phi(z,t)$, with a spatial resolution of about 100$\micron$. We
obtain the evolution of the local sedimentation velocity profiles,
$v(z,t)$, using an Image Correlation Velocimetry (ICV)
algorithm~\cite{TokumaruExpInFluids1995} that we apply to
rectangular Regions Of Interest (ROIs) of the CCD images
corresponding to the width of the sample and a height $\Delta z =
0.5~\mathrm{mm}$. To quantify the microscopic dynamics, we measure
space and time-resolved intensity correlation functions
(ICFs)\cite{DuriPRL2009}, $$g_2(z,t,\tau)-1 =
\frac{<I_p(t)I_p(t+\tau)>_z}{<I_p(t)>_z<I_p(t+\tau)>_z}-1\,,$$ where
$<\cdot \cdot \cdot >_z$ is an average over all pixels belonging to
a ROI at height $z$. Using a mixed spatio-temporal correlation
method to be described elsewhere, we correct the ICFs for the drift
due to sedimentation, so that $g_2-1$ measures the average
microscopic particle motion between times $t$ and $t +\tau$ in the
co-sedimenting frame where $v=0$. Because we detect the depolarized
scattered light, $g_2-1$ is proportional to the square of the self
part of the dynamic structure factor, $f_s(z,\mathbf{q},t,\tau) =
N^{-1} \sum_j \exp[-i\mathbf{q} \cdot (\mathbf{r}_j(t+\tau) -
\mathbf{r}_j(t))]$, where the sum is over the $N$ particles in a ROI
at height $z$ and $\mathbf{q}$ is the scattering vector. A full
decay of the ICF indicates particle rearrangements over a
lengthscale $q^{-1} = 0.66R$. Thus, we are able to measure
simultaneously and with both temporal and spatial resolution the
local volume fraction, sedimentation velocity and microscopic
relaxation dynamics, in contrast to previous works were only the
total gel height,
$h(t)$~\cite{AllainPRL1995,SenisPRE1997,DerecPRE2003,ManleyPRL2005Sedimentation,CondreJSTAT2007},
or at most $\phi(z,t)$~\cite{StarrsJPCM2002,lietor09} could be
measured.

\begin{figure}
\includegraphics[width=8.5cm]{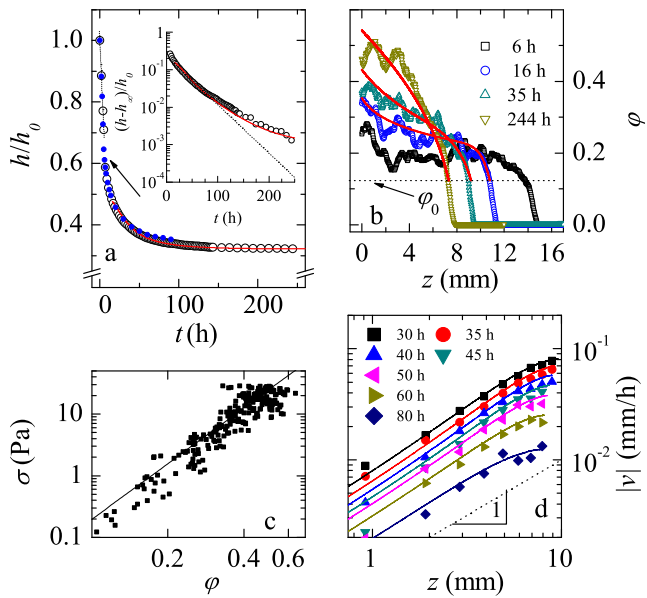}
\caption{(color online) (a) Temporal evolution of the normalized
height of the gel, $h/h_0$, for cells of section $3\times
3~\mathrm{mm}^2$ (open circles) and $5 \times 5~\mathrm{mm}^2$ (blue
dots). The arrow shows the approximate time when the gel is formed.
The solid line is the prediction of the poroelastic model,
Eq.~(\ref{eqn:SimuModel}). Inset: reduced height vs. $t$ in a
semilog plot. The dotted line is an exponential fit to the initial
decay of the reduced height, the line is the fit of the poroelastic
model. (b) Concentration profiles for various times, as shown by the
labels. The lines are fits of the poroelastic model. (c)
$\sigma(\phi)$ obtained experimentally by integrating the asymptotic
concentration profile ($t = 244$ h). The line is the power law
$\sigma [\mathrm{Pa}] = 197 \phi^3$. 
(d) Double logarithmic plot of the velocity profiles $v(z)$
(symbols, labeled by time), together with predictions from the
poroelastic model (solid lines).} \label{fig:macro}
\end{figure}

Figure~\ref{fig:macro} illustrates the time dependence of
$h(t)$, $\phi(z,t)$, and $v(z,t)$, where $z=0$ is the
cell bottom. Up to $t \approx 6~\mathrm{h}$, the sample is partitioned in a
supernatant with $\phi \approx 0$ at the top, an intermediate region with $\phi
= \phi_0$, and a denser cake that builds up from the bottom. Rapid
fluctuations of the scattered intensity indicate that the
intermediate region is formed by disconnected falling clusters,
while fluctuations are much slower in the bottom part,
revealing that a gel has formed. After $\approx6$ h, a relatively loose gel
($\phi \lesssim 0.25$, Fig.~\ref{fig:macro}b) 
occupies the entire colloid-rich phase. We follow the slow
compaction of the gel over more than 10 days, during which its
height relaxes asymptotically towards a plateau $h_{\infty} \approx
7.3$ mm. While the time evolution of $h(t)$ may be fitted by an
exponential decay at short
times~\cite{ManleyPRL2005Sedimentation,CondreJSTAT2007,KimPRL2007},
deviations are clearly observed over days (inset of
Fig.~\ref{fig:macro}a), suggesting that sedimentation is controlled
by a distribution of relaxation times, well
reproduced~\cite{Usher2006} by the poroelastic model discussed
below. The total strain of the gel is about $50\%$, such that the
gel sediments at an extremely small average strain rate of order
$\dot \epsilon \approx 10^{-6} {\rm s}^{-1}$. As shown in
Fig.~\ref{fig:macro}b, this very slow macroscopic deformation
corresponds to a nontrivial inhomogeneous evolution of the local
volume fraction $\phi(z,t)$: $\phi(z=h(t), t) \approx \phi_0$ at all
times, while at the bottom $\phi$
increases up to $\phi(z=0) \approx 0.5$.

We now turn to the behavior of the local sedimentation
velocity, a quantity not accessible in previous work.
Surprisingly, the large volume fraction inhomogeneity
has no effect on $v(z,t)$, which grows
linearly with $z$ at all times except for the uppermost layer,
see Fig.~\ref{fig:macro}d.  This
behavior suggests to characterize the sedimentation with
a single time-dependent parameter, the strain rate $\dot{\epsilon} = \mathrm{d}v/\mathrm{d}z$, a remarkable, but highly nontrivial
simplification.

Our detailed set of measurements allows us to perform a rigorous
quantitative test of the modeling of gels as a ``poroelastic''
medium pioneered by Buscall and White~\cite{BuscallFaradyTrans1987}
and used in several
studies~\cite{ManleyPRL2005Sedimentation,CondreJSTAT2007,KimPRL2007,lietor09}.
The modeling simplifies in the absence of wall friction. Our
velocimetry analysis indicates that there is no horizontal component
of the displacement resulting from the compression of the gel, i.e.
that the (effective) Poisson ratio of the gel is negligible, a
remarkable property that was discussed
before~\cite{ManleyPRL2005Sedimentation} but could not be measured
experimentally. Accordingly, no significant fraction of the
gravitational stress is redirected on the walls, making wall
friction vanish, as suggested also by the observation that changing
the cell section from $3 \times 3~\mathrm{mm}^2$ to $5 \times
5~\mathrm{mm}^2$ does not change the evolution of $h$, $\phi$, $v$
nor that of the microscopic dynamics discussed below. Neglecting
inertial terms, by balancing the gravitational stress on a gel slice
by the sum of the viscous stress, due to the flow of the suspending
fluid through the network, and the restoring stress resulting from
the gel response to a deformation, one
gets~\cite{BuscallFaradyTrans1987,buscall90}:
\begin{equation}
\frac{\partial\phi}{\partial t}=\frac{\partial}{\partial z}
\left[\frac{\phi\kappa(\phi)}{\eta}\left(\Delta\rho\phi
g+\frac{K(\phi)}{\phi}\frac{\partial\phi}{\partial z}\right)\right],
\label{eqn:SimuModel}
\end{equation}
where $K(\phi) = \phi \partial\sigma/\partial \phi$ is the effective
compressional modulus in response to an applied stress $\sigma$, and
$\kappa(\phi)$ is the gel permeability, a constant on the order of the
typical pore cross section that relates the solvent flow rate to the
pressure drop in Darcy's law
; $\eta$ and $g$ are the
solvent viscosity and the gravitational acceleration, respectively.
The boundary conditions are
$\phi(z=h,t)=\phi_0$ and zero mass flux at the cell bottom. The
initial conditions are obtained from the concentration and velocity
profiles measured experimentally at $t=16~\mathrm{h}$.

The solution of Eq.~(\ref{eqn:SimuModel}) depends crucially
on the material properties via $\kappa(\phi)$ and $K(\phi)$.
The compressional modulus may be estimated from
the concentration profile at large $t$, when $v \approx 0$~\cite{Buzzaccaro,KimPRL2007}.
For $t \rightarrow \infty$, the viscous stress vanishes and the
restoring stress $\sigma[\phi(z)]$ at a given height balances the
total weight per unit area of the particles lying above. Thus, a numerical
integration of the equilibrium profile along $z$ yields
$\sigma(\phi)$~\cite{Buzzaccaro,KimPRL2007}. As shown in Fig.~\ref{fig:macro}c, this yield stress is well fitted by $\sigma = a \phi^{3 \pm 0.3}$, where the exponent is close to that measured for the $\varphi$ dependence of the linear elastic shear modulus of colloidal gels with short-ranged interactions~\cite{PhysRevLett.82.1064}.

Various forms have been proposed in the past for the permeability
$\kappa(\phi)$, but no stringent tests on the validity of the
competing expressions could be performed, owing to the lack of
detailed measurements of the sedimentation kinetics. Expressions
assuming a fractal morphology of the
gel~\cite{ManleyPRL2005Sedimentation} fail to reproduce our data,
presumably because the volume fraction of our gel is too large for a
fully fractal structure to develop. Instead, we find that
$\kappa(\phi) = \kappa_0 \frac{(1-\phi)^m}{\phi}$, as in
Ref.~\cite{richardson54}, reproduces very well the behavior of
$h(t)$, see Fig.~\ref{fig:macro}a. The fitting parameters are
$\kappa_0 = 6.78 \times 10^{-2}~\mu\mathrm{m}^2$, corresponding to a
typical pore size in the range $0.4R - 5.9R$ (depending on $\phi$),
and $m = 7$, slightly higher that $m = 5.5 - 6.5$ typically reported
for hard spheres~\cite{richardson54,buscall90}. With $\kappa_0$ and
$m$ fixed by the fit to $h(t)$, Eq.~(\ref{eqn:SimuModel}) is solved
to obtain the temporal evolution of the full concentration and
velocity profiles. Remarkably, we find that the poroelastic model
captures very well --with no further adjustable parameters-- both
$\phi(z,t)$ and $v(z,t)$, as shown in Figs.~\ref{fig:macro}b and 1d,
demonstrating its efficient modeling of the sedimentation kinetics
of our gel at both macroscopic and mesoscopic scales, and suggesting
that the sedimentation occurs slow enough for the effective
compressional modulus to approach its asymptotic limit at $\edot
\rightarrow 0$.

Although very successful, the poroelastic model provides no insight
on the microscopic dynamics and its relation with the macroscopic
deformation, a key issue unexplored so far. Given that $K(\phi)$
appears to be unaffected by the settling kinetics, one wonders
whether the microscopic dynamics of the gel is also independent of
sedimentation. Indeed, it has been argued that gravitational
compaction provides a convenient way to slowly change $\phi$ in
order to probe the equilibrium dynamics of attractive colloidal
systems as a function of particle concentration~\cite{GaoPRL2007},
implicitly assuming that microscopic dynamics and sedimentation
decouple.

\begin{figure}
\includegraphics[width=8.5cm]{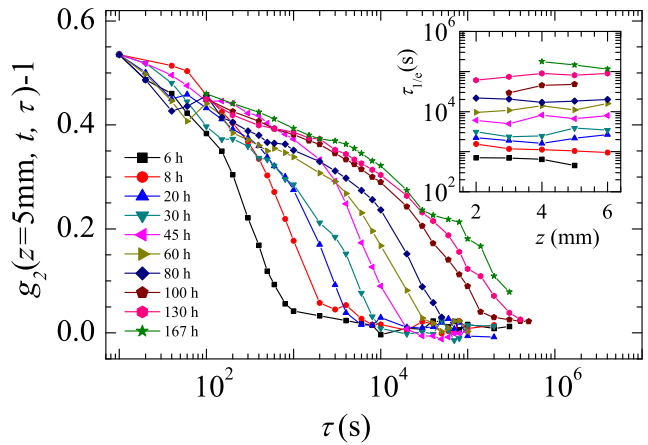}
\caption{(color online) Main panel: intensity autocorrelation function
measured at different sedimentation
times as indicated by the labels, for a cell of section $3 \times
3~\mathrm{mm}^2$. To avoid overcrowding the graph, we plot only data
for a representative height, $z = 5$~mm. Inset: relaxation time $\taue$ as a function of height $z$ for
various times (same symbols as in the main plot). For all times,
$\taue$ does not depend on $z$.}\label{fig:micro1}
\end{figure}

\begin{figure}
\includegraphics[width=8.5cm]{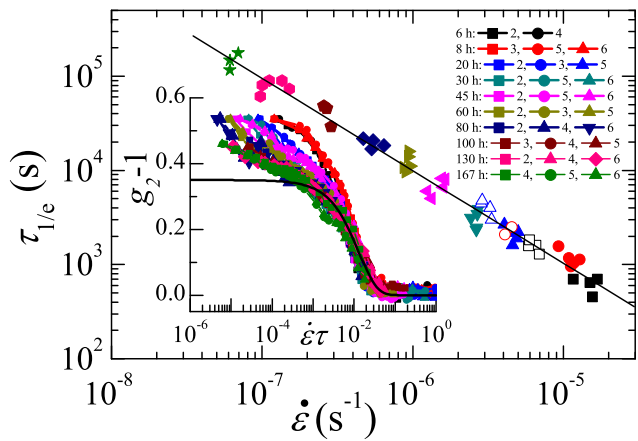}
\caption{(color online) Main panel: double logarithmic plot of
$\taue$ \emph{vs.} $\edot$. Solid line: power law fit yielding
$\taue \sim \edot^{-0.98\pm0.02}$. The solid symbols are as in
Fig.~\ref{fig:micro1}, the open symbols are data for a cell of
section $5 \times 5~\mathrm{mm}^2$ and time $t = 18, 25$, and $30$ h
for squares, circles and triangles, respectively. Inset: scaling of
$g_2-1$ measured at various heights and times (see labels, where height is in mm), when
plotted against $ \edot \tau$. The line is the function
$0.35\exp(-75\edot \tau)$.}\label{fig:micro2}
\end{figure}

To investigate the link between microscopic rearrangements and the
macroscopic gel compaction, we examine the intensity correlation
functions $g_2(z,t,\tau)-1$. Figure~\ref{fig:micro1} shows $g_2-1$
for several heights and times during the gel sedimentation.
At all $t$, a full decay of $g_2-1$ is observed, indicating that the
microstructure is continuously changing. At large $t$, an
intermediate plateau becomes increasingly visible and $g_2-1$
exhibits a two-step decay. We report in Fig.~\ref{fig:micro1}b the
decay time of the final relaxation of the ICF, $\taue$, defined by
$g_2(\taue) -1 = \mathrm{e}^{-1}[g_2(\tau= 1~\mathrm{s}) -1]$. At
any given time, $\taue$ is nearly independent of $z$. Thus, the
microscopic dynamics is fairly homogeneous across the sample, in
striking contrast with the highly inhomogeneous concentration
profiles reported in Fig.~\ref{fig:macro}b. While volume fraction
usually controls the spontaneous dynamics occurring in attractive
gels and glasses~\cite{CipellettiJPCM2005,ZaccarelliJPCM2007}, we
conclude that $\phi$ plays no such direct role in our system.
Similarly, $\taue$ is independent of $v\mathrm{_{rel}} =
v/(1-\varphi)$, the fluid velocity relative to the particles,
because $v\mathrm{_{rel}}$ grows with $z$ (data not shown) while
$\taue$ is $z$-independent. This rules out the
possibility~\cite{AllainPRL1995} that the dynamics is due to
rearrangements induced by the flow of solvent through the gel
structure.
This is further substantiated by the fact that the strain rate
associated with the gel deformation ($\edot \lesssim
10^{-5}~\mathrm{s}^{-1}$, see Fig. ~\ref{fig:micro2}) is at least
two orders of magnitude smaller than that sustained without
flow-induced damage by gels in typical linear rheology experiments.
Finally, the decay of $g_2-1$ cannot be due to the relative motion
of particles imposed by the affine vertical compression of the gel,
since the ICF is sensitive only to the horizontal component of the
displacement, parallel to $\mathbf{q}$.

Instead, we suggest that structural reorganization is in fact a
highly nonlinear process induced by the very small, but finite,
strain rate $\dot\epsilon$ of the sedimenting gel. In
colloidal~\cite{besseling07}, polymeric~\cite{lee09} and
molecular~\cite{BerthierBarrat2002} glasses under stress, the
microscopic dynamics is unaffected by the flow when the inverse
strain rate is much larger than the equilibrium structural
relaxation time. Because the latter can become very large in
amorphous materials, very small rates can enhance particle mobility,
to the point where the macroscopic flow rate controls the
microscopic relaxation time. To test this hypothesis, we plot in
Fig.~\ref{fig:micro2} $\taue$ \emph{vs.} the measured strain rate
$\edot$. The data are very well fitted by a power law, $\taue \sim
\edot^{-0.98 \pm 0.02}$ over more than two orders of magnitude,
establishing a direct relationship between microscopic structural
relaxation and macroscopic flow, $\taue \sim \edot^{-1}$, analogous
to that observed in strongly sheared glassy materials. As a further
test, we plot in the inset of Fig.~\ref{fig:micro2} $g_2-1$ against
rescaled time $ \edot \tau$.
Although the initial decay of $g_2-1$
varies with $t$, the second step of the relaxation of the ICFs
measured at different heights and times collapses reasonably well
onto a master curve (line). Note that structural rearrangement due
to plastic events occurs for a cumulated strain $\edot \taue $ of
the order of $ 1\%$, comparable to typical yield strains measured in
colloidal systems~\cite{besseling07} or molecular glass
formers~\cite{RigglemanPRL2007,Miyazaki2004}.

Our experiments provide direct evidence that in a colloidal gel
sedimenting over several days, the small external perturbation
induced by gravitational forces has dramatic consequences at the
particle scale, where structural relaxation continuously occurs at a
rate imposed by the macroscopic deformation of the sample, in
striking analogy with the enhancement of particle mobility observed
in deformed repulsive colloidal glasses~\cite{besseling07} and
polymeric glasses~\cite{lee09}. Given the intricate and non-linear
nature of the microscopic dynamics, it is all the more remarkable
that the simple macroscopic poroelastic model widely used in
sedimentation and filtration problems can successfully describe the
sedimentation kinetics of our gel to a remarkable accuracy, by
specifying only a couple of effective mechanical properties of the
material.

\acknowledgments We gratefully acknowledge funding from R\'{e}gion
Languedoc Roussillon, ANR ``Dynhet'', CNES, Institut Universitaire
de France, and the Italian Ministry of University and Research
(MIUR) (PRIN 2008). We thank Solvay-Solexis (Bollate, Italy) for
providing us with the colloidal particles.


\begin{thebibliography}{28}
\expandafter\ifx\csname natexlab\endcsname\relax\def\natexlab#1{#1}\fi
\expandafter\ifx\csname bibnamefont\endcsname\relax
  \def\bibnamefont#1{#1}\fi
\expandafter\ifx\csname bibfnamefont\endcsname\relax
  \def\bibfnamefont#1{#1}\fi
\expandafter\ifx\csname citenamefont\endcsname\relax
  \def\citenamefont#1{#1}\fi
\expandafter\ifx\csname url\endcsname\relax
  \def\url#1{\texttt{#1}}\fi
\expandafter\ifx\csname urlprefix\endcsname\relax\def\urlprefix{URL }\fi
\providecommand{\bibinfo}[2]{#2}
\providecommand{\eprint}[2][]{\url{#2}}

\bibitem[{\citenamefont{Cipelletti and Ramos}(2005)}]{CipellettiJPCM2005}
\bibinfo{author}{\bibfnamefont{L.}~\bibnamefont{Cipelletti}} \bibnamefont{and}
  \bibinfo{author}{\bibfnamefont{L.}~\bibnamefont{Ramos}}, \bibinfo{journal}{J.
  Phys.: Condens. Matter} \textbf{\bibinfo{volume}{17}}, \bibinfo{pages}{R253}
  (\bibinfo{year}{2005}).

\bibitem[{\citenamefont{Zaccarelli}(2007)}]{ZaccarelliJPCM2007}
\bibinfo{author}{\bibfnamefont{E.}~\bibnamefont{Zaccarelli}},
  \bibinfo{journal}{J. Phys.: Condens. Matter} \textbf{\bibinfo{volume}{19}},
  \bibinfo{pages}{323101} (\bibinfo{year}{2007}).

\bibitem[{\citenamefont{Royall et~al.}(2008)\citenamefont{Royall, Williams,
  Ohtsuka, and Tanaka}}]{royall08}
\bibinfo{author}{\bibfnamefont{C.~P.} \bibnamefont{Royall}},
  \bibinfo{author}{\bibfnamefont{S.~R.} \bibnamefont{Williams}},
  \bibinfo{author}{\bibfnamefont{T.}~\bibnamefont{Ohtsuka}}, \bibnamefont{and}
  \bibinfo{author}{\bibfnamefont{H.}~\bibnamefont{Tanaka}},
  \bibinfo{journal}{Nature Materials} \textbf{\bibinfo{volume}{7}},
  \bibinfo{pages}{556} (\bibinfo{year}{2008}).

\bibitem[{\citenamefont{Kasza et~al.}(2007)\citenamefont{Kasza, Rowat, Liu,
  Angelini, Brangwynne, Koenderink, and Weitz}}]{kasza07}
\bibinfo{author}{\bibfnamefont{K.~E.} \bibnamefont{Kasza}},
  \bibinfo{author}{\bibfnamefont{A.~C.} \bibnamefont{Rowat}},
  \bibinfo{author}{\bibfnamefont{J.~Y.} \bibnamefont{Liu}},
  \bibinfo{author}{\bibfnamefont{T.~E.} \bibnamefont{Angelini}},
  \bibinfo{author}{\bibfnamefont{C.~P.} \bibnamefont{Brangwynne}},
  \bibinfo{author}{\bibfnamefont{G.~H.} \bibnamefont{Koenderink}},
  \bibnamefont{and} \bibinfo{author}{\bibfnamefont{D.~A.} \bibnamefont{Weitz}},
  \bibinfo{journal}{Curr. Opin. Cell Biol.} \textbf{\bibinfo{volume}{19}},
  \bibinfo{pages}{101} (\bibinfo{year}{2007}).

\bibitem[{\citenamefont{Buscall and White}(1987)}]{BuscallFaradyTrans1987}
\bibinfo{author}{\bibfnamefont{R.}~\bibnamefont{Buscall}} \bibnamefont{and}
  \bibinfo{author}{\bibfnamefont{L.~R.} \bibnamefont{White}},
  \bibinfo{journal}{J. Chem. Soc., Faraday Trans.}
  \textbf{\bibinfo{volume}{83}}, \bibinfo{pages}{873} (\bibinfo{year}{1987}),
  \bibinfo{note}{part 3}.

\bibitem[{\citenamefont{Allain et~al.}(1995)\citenamefont{Allain, Cloitre, and
  Wafra}}]{AllainPRL1995}
\bibinfo{author}{\bibfnamefont{C.}~\bibnamefont{Allain}},
  \bibinfo{author}{\bibfnamefont{M.}~\bibnamefont{Cloitre}}, \bibnamefont{and}
  \bibinfo{author}{\bibfnamefont{M.}~\bibnamefont{Wafra}},
  \bibinfo{journal}{Phys. Rev. Lett.} \textbf{\bibinfo{volume}{74}},
  \bibinfo{pages}{1478} (\bibinfo{year}{1995}).

\bibitem[{\citenamefont{Senis and Allain}(1997)}]{SenisPRE1997}
\bibinfo{author}{\bibfnamefont{D.}~\bibnamefont{Senis}} \bibnamefont{and}
  \bibinfo{author}{\bibfnamefont{C.}~\bibnamefont{Allain}},
  \bibinfo{journal}{Phys. Rev. E} \textbf{\bibinfo{volume}{55}},
  \bibinfo{pages}{7797} (\bibinfo{year}{1997}).

\bibitem[{\citenamefont{{Poon \emph{et
  al.}}}(1999)}]{PoonFaradayDiscussion1999}
\bibinfo{author}{\bibfnamefont{W.~C.~K.} \bibnamefont{{Poon \emph{et al.}}}},
  \bibinfo{journal}{Faraday Discuss.} \textbf{\bibinfo{volume}{112}},
  \bibinfo{pages}{143} (\bibinfo{year}{1999}).

\bibitem[{\citenamefont{{Stars \textit{et al.}}}(2002)}]{StarrsJPCM2002}
\bibinfo{author}{\bibfnamefont{L.}~\bibnamefont{{Stars \textit{et al.}}}},
  \bibinfo{journal}{J. Phys.: Condens. Matter} \textbf{\bibinfo{volume}{14}},
  \bibinfo{pages}{2485} (\bibinfo{year}{2002}).

\bibitem[{\citenamefont{Derec et~al.}(2003)\citenamefont{Derec, Senis, Talini,
  and Allain}}]{DerecPRE2003}
\bibinfo{author}{\bibfnamefont{C.}~\bibnamefont{Derec}},
  \bibinfo{author}{\bibfnamefont{D.}~\bibnamefont{Senis}},
  \bibinfo{author}{\bibfnamefont{L.}~\bibnamefont{Talini}}, \bibnamefont{and}
  \bibinfo{author}{\bibfnamefont{C.}~\bibnamefont{Allain}},
  \bibinfo{journal}{Phys. Rev. E} \textbf{\bibinfo{volume}{67}},
  \bibinfo{pages}{062401} (\bibinfo{year}{2003}).

\bibitem[{\citenamefont{Manley et~al.}(2005)\citenamefont{Manley, Skotheim,
  Mahadevan, and Weitz}}]{ManleyPRL2005Sedimentation}
\bibinfo{author}{\bibfnamefont{S.}~\bibnamefont{Manley}},
  \bibinfo{author}{\bibfnamefont{J.~M.} \bibnamefont{Skotheim}},
  \bibinfo{author}{\bibfnamefont{L.}~\bibnamefont{Mahadevan}},
  \bibnamefont{and} \bibinfo{author}{\bibfnamefont{D.~A.} \bibnamefont{Weitz}},
  \bibinfo{journal}{Phys. Rev. Lett.} \textbf{\bibinfo{volume}{94}},
  \bibinfo{pages}{218302} (\bibinfo{year}{2005}).

\bibitem[{\citenamefont{Condre et~al.}(2007)\citenamefont{Condre, Ligoure, and
  Cipelletti}}]{CondreJSTAT2007}
\bibinfo{author}{\bibfnamefont{J.~M.} \bibnamefont{Condre}},
  \bibinfo{author}{\bibfnamefont{C.}~\bibnamefont{Ligoure}}, \bibnamefont{and}
  \bibinfo{author}{\bibfnamefont{L.}~\bibnamefont{Cipelletti}},
  \bibinfo{journal}{J Stat Mech - Theory and Experiments} p.
  \bibinfo{pages}{P02010} (\bibinfo{year}{2007}).

\bibitem[{\citenamefont{Kim et~al.}(2007)\citenamefont{Kim, Liu, Kuhnle, Hess,
  Viereck, Danner, Mahadevan, and Weitz}}]{KimPRL2007}
\bibinfo{author}{\bibfnamefont{C.}~\bibnamefont{Kim}},
  \bibinfo{author}{\bibfnamefont{Y.}~\bibnamefont{Liu}},
  \bibinfo{author}{\bibfnamefont{A.}~\bibnamefont{Kuhnle}},
  \bibinfo{author}{\bibfnamefont{S.}~\bibnamefont{Hess}},
  \bibinfo{author}{\bibfnamefont{S.}~\bibnamefont{Viereck}},
  \bibinfo{author}{\bibfnamefont{T.}~\bibnamefont{Danner}},
  \bibinfo{author}{\bibfnamefont{L.}~\bibnamefont{Mahadevan}},
  \bibnamefont{and} \bibinfo{author}{\bibfnamefont{D.~A.} \bibnamefont{Weitz}},
  \bibinfo{journal}{Phys. Rev. Lett.} \textbf{\bibinfo{volume}{99}},
  \bibinfo{pages}{028303} (\bibinfo{year}{2007}).

\bibitem[{\citenamefont{Lee et~al.}(2009)\citenamefont{Lee, Paeng, Swallen, and
  Ediger}}]{lee09}
\bibinfo{author}{\bibfnamefont{H.~N.} \bibnamefont{Lee}},
  \bibinfo{author}{\bibfnamefont{K.}~\bibnamefont{Paeng}},
  \bibinfo{author}{\bibfnamefont{S.~F.} \bibnamefont{Swallen}},
  \bibnamefont{and} \bibinfo{author}{\bibfnamefont{M.~D.}
  \bibnamefont{Ediger}}, \bibinfo{journal}{Science}
  \textbf{\bibinfo{volume}{323}}, \bibinfo{pages}{231} (\bibinfo{year}{2009}).

\bibitem[{\citenamefont{Besseling et~al.}(2007)\citenamefont{Besseling, Weeks,
  Schofield, and Poon}}]{besseling07}
\bibinfo{author}{\bibfnamefont{R.}~\bibnamefont{Besseling}},
  \bibinfo{author}{\bibfnamefont{E.~R.}~\bibnamefont{Weeks}},
  \bibinfo{author}{\bibfnamefont{A.~B.} \bibnamefont{Schofield}},
  \bibnamefont{and} \bibinfo{author}{\bibfnamefont{W.~C.~K.}
  \bibnamefont{Poon}}, \bibinfo{journal}{Phys. Rev. Lett.}
  \textbf{\bibinfo{volume}{99}}, \bibinfo{pages}{028301}
  (\bibinfo{year}{2007}).

\bibitem[{\citenamefont{Buzzaccaro et~al.}(2010)\citenamefont{Buzzaccaro,
  Piazza, Colombo, and Parola}}]{Buzzaccaro3}
\bibinfo{author}{\bibfnamefont{S.}~\bibnamefont{Buzzaccaro}},
  \bibinfo{author}{\bibfnamefont{R.}~\bibnamefont{Piazza}},
  \bibinfo{author}{\bibfnamefont{J.}~\bibnamefont{Colombo}}, \bibnamefont{and}
  \bibinfo{author}{\bibfnamefont{A.}~\bibnamefont{Parola}},
  \bibinfo{journal}{J. Chem. Phys.} \textbf{\bibinfo{volume}{132}},
  \bibinfo{pages}{124902} (\bibinfo{year}{2010}).

\bibitem[{\citenamefont{Buzzaccaro et~al.}(2007)\citenamefont{Buzzaccaro,
  Rusconi, and Piazza}}]{Buzzaccaro}
\bibinfo{author}{\bibfnamefont{S.}~\bibnamefont{Buzzaccaro}},
  \bibinfo{author}{\bibfnamefont{R.}~\bibnamefont{Rusconi}}, \bibnamefont{and}
  \bibinfo{author}{\bibfnamefont{R.}~\bibnamefont{Piazza}},
  \bibinfo{journal}{Phys. Rev. Lett.} \textbf{\bibinfo{volume}{99}},
  \bibinfo{pages}{098301}(\bibinfo{year}{2007}).

\bibitem{SuppMat} See EPAPS Document No. XXXXXXXXXXXX for a
supplementary figure and sample details. For more information
on EPAPS, see http:/\/www.aip.org/pubservs/epaps.html.



\bibitem[{\citenamefont{Duri et~al.}(2009)\citenamefont{Duri, Sessoms, Trappe,
  and Cipelletti}}]{DuriPRL2009}
\bibinfo{author}{\bibfnamefont{A.}~\bibnamefont{Duri}},
  \bibinfo{author}{\bibfnamefont{D.~A.} \bibnamefont{Sessoms}},
  \bibinfo{author}{\bibfnamefont{V.}~\bibnamefont{Trappe}}, \bibnamefont{and}
  \bibinfo{author}{\bibfnamefont{L.}~\bibnamefont{Cipelletti}},
  \bibinfo{journal}{Phys. Rev. Lett.} \textbf{\bibinfo{volume}{102}},
  \bibinfo{pages}{085702} (\bibinfo{year}{2009}).


\bibitem[{\citenamefont{Tokumaru and
  Dimotakis}(1995)}]{TokumaruExpInFluids1995}
\bibinfo{author}{\bibfnamefont{P.~T.} \bibnamefont{Tokumaru}} \bibnamefont{and}
  \bibinfo{author}{\bibfnamefont{P.~E.} \bibnamefont{Dimotakis}},
  \bibinfo{journal}{Exp. Fluids} \textbf{\bibinfo{volume}{19}},
  \bibinfo{pages}{1} (\bibinfo{year}{1995}).

\bibitem[{\citenamefont{Lietor-Santos et~al.}(2009)\citenamefont{Lietor-Santos,
  Kim, Lu, Fernandez-Nieves, and Weitz}}]{lietor09}
\bibinfo{author}{\bibfnamefont{J.~J.} \bibnamefont{Lietor-Santos}},
  \bibinfo{author}{\bibfnamefont{C.}~\bibnamefont{Kim}},
  \bibinfo{author}{\bibfnamefont{P.}~\bibnamefont{Lu}},
  \bibinfo{author}{\bibfnamefont{A.}~\bibnamefont{Fernandez-Nieves}},
  \bibnamefont{and} \bibinfo{author}{\bibfnamefont{D.}~\bibnamefont{Weitz}},
  \bibinfo{journal}{Eur. Phys. J. E} \textbf{\bibinfo{volume}{28}},
  \bibinfo{pages}{159} (\bibinfo{year}{2009}).

\bibitem[{\citenamefont{Usher et~al.}(2006)\citenamefont{Usher, Scales, and
  White}}]{Usher2006}
\bibinfo{author}{\bibfnamefont{S.~P.} \bibnamefont{Usher}},
  \bibinfo{author}{\bibfnamefont{P.~J.} \bibnamefont{Scales}},
  \bibnamefont{and} \bibinfo{author}{\bibfnamefont{L.~R.} \bibnamefont{White}},
  \bibinfo{journal}{AIChe J.} \textbf{\bibinfo{volume}{52}},
  \bibinfo{pages}{986} (\bibinfo{year}{2006}).

\bibitem[{\citenamefont{Buscall}(1990)}]{buscall90}
\bibinfo{author}{\bibfnamefont{R.}~\bibnamefont{Buscall}},
  \bibinfo{journal}{Colloids and Surfaces} \textbf{\bibinfo{volume}{43}},
  \bibinfo{pages}{33} (\bibinfo{year}{1990}).

\bibitem[{\citenamefont{Gisler et~al.}(1999)\citenamefont{Gisler, Ball, and
  Weitz}}]{PhysRevLett.82.1064}
\bibinfo{author}{\bibfnamefont{T.}~\bibnamefont{Gisler}},
  \bibinfo{author}{\bibfnamefont{R.~C.} \bibnamefont{Ball}}, \bibnamefont{and}
  \bibinfo{author}{\bibfnamefont{D.~A.} \bibnamefont{Weitz}},
  \bibinfo{journal}{Phys. Rev. Lett.} \textbf{\bibinfo{volume}{82}},
  \bibinfo{pages}{1064} (\bibinfo{year}{1999}).

\bibitem[{\citenamefont{Richardson and Zaki}(1954)}]{richardson54}
\bibinfo{author}{\bibfnamefont{J.~F.} \bibnamefont{Richardson}}
  \bibnamefont{and} \bibinfo{author}{\bibfnamefont{W.~N.} \bibnamefont{Zaki}},
  \bibinfo{journal}{Trans. Inst. Chem. Eng} \textbf{\bibinfo{volume}{32}},
  \bibinfo{pages}{35} (\bibinfo{year}{1954}).

\bibitem[{\citenamefont{Gao and Kilfoil}(2007)}]{GaoPRL2007}
\bibinfo{author}{\bibfnamefont{Y.}~\bibnamefont{Gao}} \bibnamefont{and}
  \bibinfo{author}{\bibfnamefont{M.~L.} \bibnamefont{Kilfoil}},
  \bibinfo{journal}{Phys. Rev. Lett.} \textbf{\bibinfo{volume}{99}},
  \bibinfo{pages}{078301} (\bibinfo{year}{2007}).

\bibitem[{\citenamefont{Berthier and Barrat}(2002)}]{BerthierBarrat2002}
\bibinfo{author}{\bibfnamefont{L.}~\bibnamefont{Berthier}} \bibnamefont{and}
  \bibinfo{author}{\bibfnamefont{J.-L.} \bibnamefont{Barrat}},
  \bibinfo{journal}{J. Chem. Phys.} \textbf{\bibinfo{volume}{116}},
  \bibinfo{pages}{6228} (\bibinfo{year}{2002}).

\bibitem[{\citenamefont{Riggleman et~al.}(2007)\citenamefont{Riggleman, Lee,
  Ediger, and De~Pablo}}]{RigglemanPRL2007}
\bibinfo{author}{\bibfnamefont{R.~A.} \bibnamefont{Riggleman}},
  \bibinfo{author}{\bibfnamefont{H.~N.} \bibnamefont{Lee}},
  \bibinfo{author}{\bibfnamefont{M.~D.} \bibnamefont{Ediger}},
  \bibnamefont{and} \bibinfo{author}{\bibfnamefont{J.~J.}
  \bibnamefont{de~Pablo}}, \bibinfo{journal}{Phys. Rev. Lett.}
  \textbf{\bibinfo{volume}{99}}, \bibinfo{pages}{215501}
  (\bibinfo{year}{2007}).

\bibitem[{\citenamefont{Miyazaki et~al.}(2004)\citenamefont{Miyazaki, Reichman,
  and Yamamoto}}]{Miyazaki2004}
\bibinfo{author}{\bibfnamefont{K.}~\bibnamefont{Miyazaki}},
  \bibinfo{author}{\bibfnamefont{D.~R.} \bibnamefont{Reichman}},
  \bibnamefont{and} \bibinfo{author}{\bibfnamefont{R.}~\bibnamefont{Yamamoto}},
  \bibinfo{journal}{Phys. Rev. E} \textbf{\bibinfo{volume}{70}},
  \bibinfo{pages}{011501} (\bibinfo{year}{2004}).

\end{thebibliography}

\end{document}